\documentstyle[12pt]{article}
\newlength{\extraspace}
\newlength{\extraspaces}
\newcommand{\ba}{\begin{eqnarray}
\addtolength{\abovedisplayskip}{\extraspaces}
\addtolength{\belowdisplayskip}{\extraspaces}
\addtolength{\abovedisplayshortskip}{\extraspace}
\addtolength{\belowdisplayshortskip}{\extraspace}}
\newcommand{\ea}{\end{eqnarray}}

\newcommand{\nonu}{\nonumber \\[.5mm]}
\newcommand{\A}{&\!\!\!}

\begin{document}

\thispagestyle{empty}

\hfill \parbox{3.5cm}{SIT-LP-12/04}
\vspace*{1cm}
\vspace*{1cm}
\begin{center}
{\large \bf Nonliner supersymmetric general relativity \\ 
and unity of nature \\
-Is the real shape of nature unstable spacetime?- }
\footnote{Based upon the talk given by K. Shima at the XIXth Max Born Symposium,  
September 28-October 1, 2004, Wroclaw University, Poland}  \\[20mm]  
{\bf Kazunari SHIMA and Motomu TSUDA} \\[2mm]
{\em Laboratory of Physics,  Saitama Institute of Technology}\footnote{e-mail: shima@sit.ac.jp, tsuda@sit.ac.jp}\\
{\em Okabe-machi, Saitama 369-0293, Japan}\\[2mm]
{December 2004}\\[15mm]

%\maketitle

\begin{abstract}
The Einstein-Hilbert-type action for nonlinear supersymmetric(NLSUSY) general relativity(GR) 
proposed as the fundamental action  for nature is written down explicitly 
in terms of the fundamental fields, the graviton and the Nambu-Goldstone(NG) fermion(superons). 
For comparisons the expansion of the action is carried out by using the affine connection formalism 
and the spin connection formalism. 
The linearization of NLSUSY GR is considered and carried out explicitly for the N=2 NLSUSY(Volkov-Akulov) 
model, which reproduce the equivalent renormalizable theory of the gauge vector multiplet of  N=2 LSUSY.  
Some characteristic structures including some hidden symmetries 
of the gravitational coupling of superons are manifested 
(in two dimensional space-time) with some details of the calculations. 
SGM cosmology is discussed briefly. 

\end{abstract}
\end{center}

\newpage
{\bf 1. Introduction}             \\
Supersymmetry(SUSY)\cite{wz}\cite{va}\cite{gl}is recognized as the most 
promissing gauge symmetry beyond the standard model(SM), especially for the unification 
of space-time and matter. In fact the theory of supergravity(SUGRA) 
is constructed based upon the local SUSY, which brings the breakthrough 
for the unification of space-time and matter\cite{FVF}\cite{DZ}. 
Consequently Nambu-Goldstone(N-G) fermion \cite{va}\cite{SS} would appear 
in the spontaneous SUSY breaking and plays essentially important roles 
in the unified model building.             \\
Here it is useful to distinguish the qualitative differences 
of the origins of N-G fermion. 
In O'Raifeartaigh model\cite{o}  N-G fermion stems from 
the symmetry of the dynamics(interaction) of the linear 
representation multiplet of SUSY, i.e. it corresponds 
to the coset space coordinates of G/H where G and H are realized 
on the field operators. 
While in Volkov-Akulov (V-A) model\cite{va} N-G fermion stems 
from the degrees of freedom(d.o.f.) of the symmetry (breaking) of spacetime 
G/H in terms of the (supersymmetric) geometrical arguments 
and gives the nonlinear(NL) representation of SUSY.    \\
As demonstrated in SUGRA coupled with V-A model it is rather well understood 
in the linear realization of  SUSY(L SUSY) that N-G fermion is 
converted to the longitudinal component of spin 3/2 gravitino field 
by the super-Higgs mechanism and  breaks local linear SUSY spontaneously 
by giving mass to gravitino\cite{vsdz}. 
N-G fermion degrees of freedom become unphysical in the low energy.       \\
The SM and the grand unified theory(GUT) equiptted naively with SUSY 
have revealed the remarkable features, e.g. the unification 
of the gauge couplings at about $10^{17}$, relatively 
stable proton(now almost excluded in the simple model),etc., 
but in general they possess more than one hundred arbitrary parameters and particles 
and less predictive  powers and the gravity is out of the scope. 
While, considering seriously the fact  that SUSY is 
naturally connected to spacetime symmetry, 
it may be interesting to survey other possibilities 
concerning how SUSY is realized and where N-G fermion degrees of freedom 
have gone (in the low energy).                  \\
Facing so many fundamental elementary particles 
(more than 160 for SUSY GUTs) and  arbitrary coupling parameters, 
we are tempted to suppose that they may be certain composites  and/or that 
they should be attributed to the  particular geometrical structure of spacetime. 
In fact, concerning SUSY the various types of the composite models of the elementary particles 
are proposed\cite{BV}\cite{Lki}. 
In Ref.\cite{BV}, N-G fermion is considered as the fundamental 
constituents of matter (quarks and leptons) 
and the field theoretical description of the model is attempted. 
From the viewpoints of simplicity and beauty of nature 
the unified theory should accommodate all observed  particles in a single
irreducible representation of a certain algebra(group) especially in the case of 
spacetime having a certain boundary.   
In Ref. \cite{ks1} we have found group theoretically that among the massless irreducible representations 
of all SO(N) super-Poincar\'e(SP) groups 
N=10 SP group is the only one that contains the strong-electroweak standard model(SM) 
with just three genarations of quarks and leptons, where we have decomposed 10 supercharges into  
${\underline 10 = \underline 5 + \underline 5^{*}}$ with respect to SU(5). 
Interestingly, the quantum numbers of the superon-quintet 
are the same as those of the fundamental representation 
${\underline 5}$ of the matter multiplet of SU(5) GUT\cite{gg}. 
Regarding 10 {\it supercharges} as the hypothetical fudamental spin 1/2 {\it particles(superons)}-quintet 
and anti-quintet, we have proposed the composite superon-graviton model(SGM) for nature\cite{ks1}.
In SGM, all observed elementary particles including gravity are assigned 
to a single irreducible massless representation of SO(10) 
super-Poincar\'e(SP) symmetry and reveals a remarkable potential 
for the phenomenology, e.g. they may explain naturally, 
though qualitatively at moment, 
the three-generations structure of quarks and leptons, 
the stability of proton, the origins of various mixing angles, 
CP-violation phase and the (small) Yukawa coupling constants, 
..etc\cite{ks1}\cite{sts1}.
And except graviton they are supposed to be 
the (massless) eigenstates of superons of 
SO(10) SP symmetry \cite{ks2} of space-time and matter. 
This group theoretical argument and the 
structure of the supercurrent\cite{ks0} indicate the {\it field-current(charge) identity} 
for the fundamental theory, 
i.e. N=10 NLSUSY Volkov-Akulov(VA) model~\cite{va} in curved spacetime. 
The uniqueness of N=10 among all SO(N) SP is pointed out.
The arguments are group theoretical so far.  \\
In order to obtain the fundamental action of SGM which is invariant 
at least under local GL(4,R), local Lorentz, global NL SUSY 
transformations and global SO(10), 
we have performed the similar geometrical arguments to 
Einstein genaral relativity theory(EGRT) 
in high symmetric  SGM space-time, 
where the tangent (Riemann-flat) space-time is specified 
by (the coset space coordinates corresponding to) N-G fermion 
of NL SUSY of V-A\cite{va} 
in addition to the ordinary Lorentz SO(3,1) coordinates\cite{ks1}, 
which are locally homomorphic groups\cite{ks3}. 
As shown in Ref.\cite{ks3} the SGM action 
for the unified SGM space-time is defined 
as the geometrical invariant quantity 
and is naturally the analogue of Einstein-Hilbert(E-H) action of GR 
which has the similar concise expression. 
And interestingly it may be regarded as a kind of a generalization 
of Born-Infeld action\cite{bi}. 
(The similar systematic arguments are applicable to 
spin 3/2 N-G case.\cite{st1})                           \\
In this article, after a brief review of SGM 
for the self contained arguments we write down SGM action 
in terms of the fields of graviton and superons 
in order to see some characteristic structures of our model 
and also show some details of the calculations.     \\
For the sake of the comparison the expansions are performed 
by the affine connection formalism and by the spin connection formalism, 
which are equivalent to the power series expansions 
in the universal coupling constant 
(the fundamental volume of four dimensional spacetime).            \\
Finally some hidden symmetries, the linearlization(i.e.,the derivation 
of the equivalent low energy theory) 
of SGM and a potential cosmology, 
especially the birth of the universe are mentioned briefly. 

{\bf 2. Fundamental action  NLSUSY General Relativity}       \\
A nonlinear supersymmetric general relativity theory(NLSUSY GRT) 
(i.e.,  N=1 SGM action ) is proposed.  
We extend the geometrical arguments of Einstein general relativity theory(EGRT) on Riemann spacetime 
to new  (SGM) spacetime posessing ${\it locally}$   NL SUSY d.o.f, 
i.e.   besides the ordinary SO(3,1) Minkowski coordinate $x^{a}$ 
the SL(2C) Grassman coordinates $\psi$ for the coset space ${superGL(4,R) \over GL(4,R)}$ 
turning subsequently to the NG fermion dynamical d.o.f. are attached at every curved spacetime point. 
Note that SO(3,1) and SL(2C) are locally holomorphic {\it non-compact groups for spacetime (coordinates) d.o.f.}, 
which may be analogous to SO(3) and SU(2) {\it compact groups for gauge (fields) d.o.f.} of 
't Hooft-Polyakov monopole\cite{th}\cite{p}.  
We have obtained  the following NLSUSY GRT(N=1 SGM) action\cite{ks2} of the vacuum EH-type.
\begin{equation}
L(w)=-{c^{4} \over 16{\pi}G}\vert w \vert(\Omega + \Lambda ),
\label{SGM}
\end{equation}
\begin{equation}
\vert w \vert=det{w^{a}}_{\mu}=det({e^{a}}_{\mu}+ {t^{a}}_{\mu}(\psi)),  \quad
{t^{a}}_{\mu}(\psi)={\kappa^{4}  \over 2i}(\bar{\psi}\gamma^{a}
\partial_{\mu}{\psi}
- \partial_{\mu}{\bar{\psi}}\gamma^{a}{\psi}),
\label{w}
\end{equation} 
where $w^{a}{_\mu}(x)$ is the unified vierbein of SGM spacetime, 
G is the gravitational constant, ${\kappa^{4} = ({c^{4}\Lambda \over 16{\pi}G}})^{-1}$ 
is a fundamental volume of four dimensional spacetime of VA model~\cite{va},  
and $\Lambda$ is a  ${small}$ cosmological constant related to the strength of 
the superon-vacuum coupling constant. 
Therefore SGM contains two mass scales,  ${1 \over {\sqrt G}}$(Planck scale) in the first term describing 
the curvature energy  and $\kappa \sim {\Lambda \over G}(O(1))$ in the second term describing the 
vacuum energy of SGM, which are responsible for the masss hierarchy.
$e^{a}{_\mu}$ is the ordinary vierbein of EGRT  describing the local SO(3,1) d.o.f 
and  ${t^{a}}_{\mu}(\psi)$ itself is not the vierbein but the mimic vierbein analogue composed of 
the stress-energy-momentum tensor of superons describing the local SL(2C) d.o.f..
$\Omega$ is a new scalar curvature analogous to the Ricci scalar curvature $R$ of EGRT, 
whose explicit expression is obtained  by just replacing ${e^{a}}_{\mu}(x)$  
by ${w^{a}}_{\mu}(x)$ in Ricci scalar $R$~\cite{st1}.    \\
These results can be understood intuitively by observing that 
${w^{a}}_{\mu}(x) ={e^{a}}_{\mu}+ {t^{a}}_{\mu}(\psi)$  inspired  by 
$\omega^{a}=dx^{a} + {\kappa^{4}  \over 2i}(\bar{\psi}\gamma^{a}
d{\psi}
- d{\bar{\psi}}\gamma^{a}{\psi})
\sim {w^{a}}_{\mu}dx^{\mu}$, where $\omega^{a}$ is the NLSUSY invariant differential forms of 
VA\cite{va}, is invertible, i.e.,
\begin{equation}
w^{\mu}{_a}= e^{\mu}{_a}- t{^{\mu}}_a + {t^{\mu}}_{\rho}{t^{\rho}}_a 
- t{^{\mu}}_{\sigma} t{^{\sigma}}_{\rho}   t{^{\rho}}_a 
+t{^{\mu}}_{\kappa} t{^{\kappa}}_{\sigma}t{^{\sigma}}_{\rho}t{^{\rho}}_a + \cdots, 
\label{w-inverse}
\end{equation} 
which terminates with $(t)^{4}$ and $s_{\mu\nu} \equiv w^{a}{_\mu}\eta_{ab}w^{b}{_\nu}$ and 
$s^{\mu \nu}(x) \equiv w^{\mu}{_{a}}(x) w^{{\nu}{a}}(x)$ 
are a unified vierbein and a unified metric tensor of NLSUSY GRT in SGM spacetime\cite{{ks2},{st1}}. 
It is straightforward to show 
${w_{a}}^{\mu} w_{{\mu}{b}} = \eta_{ab}$,  $s_{\mu \nu}{w_{a}}^{\mu} {w_{b}}^{\mu}= \eta_{ab}$, ..etc. 
As read out in (\ref{w}), in contrast with EGRT we must be careful with 
{\it the order of the indices of the tensor in NLSUSY GRT}, 
i.e. the first and the second index of $w$ (and $t$)  represent those of the $\gamma$-matrix 
and the derivative on $\psi$, respectively.  
It seems natural that the ordinary vierbein $e^{a}{_\mu}$ and the mimic vierbein $t^{a}{_\mu}$ of 
the stress-enery-momentum tensor of superon are alined and contribute equally 
to the unified vierbein  $w^{a}{_\mu}$, 
i.e. to the curvature (total energy) of unified SGM spacetime, 
where the  fundamental  action of NLSUSY GRT is the  vacuum (empty space)  action of EH-type.   \\
NLSUSY GR action  (\ref{SGM}) is invariant at least under the following transformations\cite{st2};
the following new NLSUSY transformation 
\begin{equation}
\delta^{NL} \psi ={1 \over \kappa^{2}} \zeta + 
i \kappa^{2} (\bar{\zeta}{\gamma}^{\rho}\psi) \partial_{\rho}\psi,
\quad
\delta^{NL} {e^{a}}_{\mu} = i \kappa^{2} (\bar{\zeta}{\gamma}^{\rho}\psi)\partial_{[\rho} {e^{a}}_{\mu]},
\label{newsusy}
\end{equation} 
where $\zeta$ is a constant spinor and  $\partial_{[\rho} {e^{a}}_{\mu]} = 
\partial_{\rho}{e^{a}}_{\mu}-\partial_{\mu}{e^{a}}_{\rho}$, \\
the following GL(4R) transformations due to (\ref{newsusy})  
\begin{equation}
\delta_{\zeta} {w^{a}}_{\mu} = \xi^{\nu} \partial_{\nu}{w^{a}}_{\mu} + \partial_{\mu} \xi^{\nu} {w^{a}}_{\nu}, 
\quad
\delta_{\zeta} s_{\mu\nu} = \xi^{\kappa} \partial_{\kappa}s_{\mu\nu} +  
\partial_{\mu} \xi^{\kappa} s_{\kappa\nu} 
+ \partial_{\nu} \xi^{\kappa} s_{\mu\kappa}, 
\label{newgl4r}
\end{equation} 
where  $\xi^{\rho}=i \kappa^{2} (\bar{\zeta}{\gamma}^{\rho}\psi)$, 
and the following local Lorentz transformation on $w{^a}_{\mu}$ 
\begin{equation}
%\eqalign{
\delta_L w{^a}_{\mu}
= \epsilon{^a}_b w{^b}_{\mu}
%}
\label{Lrw}
\end{equation}
with the local  parameter
$\epsilon_{ab} = (1/2) \epsilon_{[ab]}(x)$    
or equivalently on  $\psi$ and $e{^a}_{\mu}$
\begin{equation}
\delta_L \psi = - {i \over 2} \epsilon_{ab}
      \sigma^{ab} \psi,     \quad
\delta_L {e^{a}}_{\mu} = \epsilon{^a}_b e{^b}_{\mu}
      + {\kappa^{4} \over 4} \varepsilon^{abcd}
      \bar{\psi}\gamma_5 \gamma_d \psi
      (\partial_{\mu} \epsilon_{bc}).
\label{newlorentz}
\end{equation}
The local Lorentz transformation forms a closed algebra, for example, on $e{^a}_{\mu}$ 
\begin{equation}
[\delta_{L_{1}}, \delta_{L_{2}}] e{^a}_{\mu}
= \beta{^a}_b e{^b}_{\mu}
+ {\kappa^{4} \over 4} \varepsilon^{abcd} \bar{\psi}
\gamma_5 \gamma_d \psi
(\partial_{\mu} \beta_{bc}),
\label{comLr1/2}
\end{equation}
where $\beta_{ab}=-\beta_{ba}$ is defined by
$\beta_{ab} = \epsilon_{2ac}\epsilon{_1}{^c}_{b} -  \epsilon_{2bc}\epsilon{_1}{^c}_{a}$.
The commutators of two new NLSUSY transformations (\ref{newsusy})  on $\psi$ and  ${e^{a}}_{\mu}$ 
are GL(4R), i.e. new NLSUSY (\ref{newsusy}) is the square-root of GL(4R); 
\begin{equation}
[\delta_{\zeta_1}, \delta_{\zeta_2}] \psi
= \Xi^{\mu} \partial_{\mu} \psi,
\quad
[\delta_{\zeta_1}, \delta_{\zeta_2}] e{^a}_{\mu}
= \Xi^{\rho} \partial_{\rho} e{^a}_{\mu}
+ e{^a}_{\rho} \partial_{\mu} \Xi^{\rho},
\label{com1/2-e}
\end{equation}
where 
$\Xi^{\mu} = 2i\kappa (\bar{\zeta}_2 \gamma^{\mu} \zeta_1)
      - \xi_1^{\rho} \xi_2^{\sigma} e{_a}^{\mu}
      (\partial_{[\rho} e{^a}_{\sigma]})$.
They show the closure of the algebra. 
The ordinary local GL(4R) invariance is trivial by the construction.   
Besides these familiar and intended symmetries, 
the unified vierbein $w^{a}{_\mu}$, therefore SGM action, 
is invariant under the following local spinor translation(ST) 
with the local spinor paremeter $\theta(x)$; 
${\delta \psi=\theta}$,    
$\delta e^{a}{_\mu}=
-i \kappa^{2}( \bar\theta\gamma^{a} \partial_{\mu}\psi+\bar\psi\gamma^{a} \partial_{\mu}\theta )$. 
The commutators vanish identically.
Note that the NG fermion d.o.f. $\psi$ can be transformed(redefined) away 
neither by this local ST, 
in fact, ${w(e + \delta e, t(\psi + \delta \psi))=w(e + t(\psi), 0)=w(e,t(\psi))}$ under ${\theta(x)=-\psi(x)}$ 
as indicated ${\delta w^{a}{_\mu}(x)=0}$(an invariant quantity) 
nor by the ordinary general coordinate transformation d.o.f. 
$\delta_{GL(4R)}e^{a}{_\mu}$, for the unconstrained such d.o.f. 
are reserved for the arbitrary on-shell (gauge) condition for (surviving) $e^{a}{_\mu}$ itself.   
Otherwise, $\delta_{GL(4R)}e^{a}{_\mu}$ induces a restricted and 
pathological on-shell conditions.
Therefore NLSUSY spacetime d.o.f. is preserved.
Taking $\psi=0$ by hand makes SGM another theory(EH theory) based upon another flat space(Minkowski spacetime). 
This local spinor coordinate translation invariance is somewhat puzzling(immature) so far 
but is the origin of (or recasted as) the local spinor {\it gauge} symmetry 
of the  {\it linear SUSY  gauge field} theory (SUGRA-analogue yet to be obtained by the linearization) 
which is equivalent to SGM  and the mass generation through the super-Higgs mechanism 
in the spontaneous symmetry breakdown.
The extension to N=10, i.e. SO(10) SP is straightforward by taking ${\psi^{j}, (j=1,2,..10)}$\cite{ks2}.   \\
Now NLSUSY GRT action (\ref{SGM}) is invariant at least under the following spacetime symmetries\cite{st2}
\begin{equation}
[{\rm new \ NLSUSY}] \otimes [{\rm local\ GL(4,R)}] 
\otimes [{\rm local\ Lorentz }] \otimes [{\rm local \ ST}],  
\label{sgmspsymm}
\end{equation}
which is isomorphic to N=1 SP group of SUGRA. 
The extension to N=10, i.e. SO(10) SP is straightforward by taking ${\psi^{j}, (j=1,2,..10)}$\cite{ks2}.   \\
As for the internal summetry we just mention that  ${w^{a}{_\mu}}$ is invariant 
under the local U(1) transformation $\psi{_j} \rightarrow e^{i \lambda_{j}(x)}\psi_{j}$ 
due to the Grassman(Majorana spinor) nature $\bar{\psi_{j}}\gamma^{a}{\psi_{j}}=0$, i.e. 
(\ref{SGM}) with N-extension is invariant at least under 
\begin{equation}
[{global \ SO(N)}] \otimes [local \ U(1)]^{N}.  
\label{sgmisymm}
\end{equation}
Therefore the action (\ref{SGM}) describes the vacuum energy(everything) of the ultimate spacetime and 
is NLSUSY GRT, a nontrivial generalization of the EH action.   
It should be noticed that SGM action  (\ref{SGM}) posesses two types of flat space which are 
not equivalent, i.e. SGM-flat($w{^a}_{\mu}(x) \rightarrow {\delta}{^a}_{\mu}$)  and 
Riemann-flat($e{^a}_{\mu}(x) \rightarrow {\delta}{^a}_{\mu}$). 
As discussed later this structure plays impotant roles in the spontaneous breakdown of spacetime and 
in the cosmology of SGM (\ref{SGM}).   
The linearization of SGM action (\ref{SGM}) and identifying an equivalent, local and renormalizable 
{\it gauge} field theory is inevitable to test SGM scenario, though some characterictic predictions are 
presented~\cite{ks1,ks2} qualitatively. \\
Finally we just mention the hidden symmetries  characteristic to SGM.  
It is natural to expect that SGM action may be invariant 
under a certain exchange 
between ${e^{a}}_{\mu}$ and ${t^{a}}_{\mu}$, 
for they contribute equally to the unified SGM vierbein  ${w^{a}}_\mu$ 
as seen in (\ref{new-w}).
In fact we find, as a simple example, that ${w^{a}}_\mu$ and  
${w_{a}}^\mu$, i.e. SGM action is invariant under the following exchange  
of ${e^{a}}_{\mu}$ and ${t^{a}}_{\mu}$ (in 4 dimensional space-time).  
\ba
\A \A {e^{a}}_{\mu}  \longrightarrow  2{t^{a}}_{\mu}, 
{t^{a}}_{\mu} \longrightarrow  {e^{a}}_{\mu} -  {t^{a}}_{\mu},  \nonu
\A \A \hspace{1.5cm} 
{e_{a}}^{\mu} \longrightarrow {e_{a}}^{\mu}, 
%}
\label{e-tEX}
\ea
or in terms of the  metric it can be written as  
\ba
\A \A g_{\mu\nu} \longrightarrow 4{t^{\rho}}_{\mu}t_{\rho\nu}, 
t_{\mu\nu} \longrightarrow 2(t_{\nu\mu}-t_{\rho\mu}{t^{\rho}}_{\nu}),  \nonu
\A \A \hspace{1.5cm} 
g^{\mu\nu} \longrightarrow g^{\mu\nu}, t^{\mu\nu} \longrightarrow  g^{\mu\nu} - t^{\mu\nu}. 
%}
\label{g-tEX}
\ea
This can be generalized to the following form with two real(one complex) global prameters,         \\
\begin{equation}
\pmatrix{e{^a}_{\mu} \cr
         t{^a}_{\mu} \cr
         t{^b}_{\mu} e{_b}^{\nu} t{^a}_{\nu} \cr} 
\rightarrow \pmatrix{
            0 & 2(\alpha + 1) & -2(\alpha^2 - \beta) \cr
            1 & -(2 \alpha + 1) & 2(\alpha^2 - \beta) \cr
            1 & -(3 \alpha + 2) & 2 \alpha(2 \alpha + 1) - 3 \beta + 1 \cr} 
\ \pmatrix{e{^a}_{\mu} \cr
         t{^a}_{\mu} \cr
         t{^b}_{\mu} e{_b}^{\nu} t{^a}_{\nu} \cr}. 
\end{equation}
The physical meaning  of such   symmetries  remains to be studied.         \\   
Also SGM action has $Z_{2}$ symmetry $\psi^{j} \rightarrow -\psi^{j}$.              \\
Now  to clarify the characteristic features of SGM 
we focus on  N=1 SGM  for simplicity  without loss of generality and 
write down the action explicitly in terms of ${t^{a}}_{\mu}$(or $\psi$) 
and $g^{\mu\nu}$(or ${e^{a}}_{\mu}$). We will see that the graviton and superons(matter) are complementary 
in SGM and contribute equally to the curvature of SGM space-time. 
Contrary to its  simple expression (\ref{SGM}), it has rather complicated and rich structures.   \\
To obtain (\ref{SGM}) we require that the unified action of SGM space-time should reduce to V-A 
in the  flat space-time which is specified by $x^{a}$ and $\psi(x)$ and that 
the graviton and superons contribute equally to the unified curvature of SGM space-time. 
We have found that the unified vierbein  $w{^a}_{\mu}(x)$ and the unified metric $s_{\mu\nu}(x)$ 
of unified SGM space-time are defined  through the NL SUSY invariant differential
forms $\omega^a$ of V-A\cite{va} as follows: 
\ba
\A \A \omega^a \sim w{^a}_{\mu} dx^{\mu},
\label{om} \\
\A \A w{^a}_{\mu}(x) = e{^a}_{\mu}(x) + t{^a}_{\mu}(x),
\label{new-w}
\ea
where $e{^a}_{\mu}(x)$ is the vierbein of EGRT and $t{^a}_{\mu}(x)$
is defined by
\begin{equation}
%\eqalign{
t{^a}_{\mu}(x) = i\kappa \bar{\psi}\gamma^{a}
\partial_{\mu}{\psi},
%}
\label{t-1/2}
\end{equation}
where the first and the second indices of ${t^{a}}_{\mu}$ represent those of the $\gamma$ matrices and 
the general covariant derivatives, respectively.
We can easily obtain the inverse $w{_a}^{\mu}$ of the  vierbein
$w{^a}_{\mu}$ in the power series of $t{^a}_{\mu}$ as follows, 
which terminates with $t^4$(for 4 dimensional space-time):
\begin{equation}
%\eqalign{
w{_a}^{\mu} = e{_a}^{\mu}
- t{^{\mu}}_a + t{^{\rho}}_a t{^{\mu}}_{\rho} - t{^{\rho}}_a t{^{\sigma}}_{\rho} t{^{\mu}}_{\sigma} 
+ t{^{\rho}}_a t{^{\sigma}}_{\rho} t{^{\kappa}}_{\sigma}t{^{\mu}}_{\kappa}.
%}
\label{new-wi}
\end{equation}
Similarly a new metric tensor $s_{\mu\nu}(x)$ and its inverse $s^{\mu\nu}(x)$ 
are introduced in SGM curved space-time as follows:
\ba
\A \A s_{\mu\nu}(x) \equiv w{^a}_{\mu}(x) w_{a \nu}(x) =w{^a}_{\mu}(x) \eta_{ab} w{^b}_{ \nu}(x) \nonu
\A \A \hspace{1.5cm}
= g_{\mu\nu} + t_{\mu\nu} + t_{\nu\mu} + {t^{\rho}}_{\mu} {t_{\rho\nu}}.
\label{new-s}
\ea
and 
\ba
\A \A s^{\mu\nu}(x) \equiv w{_a}^{\mu}(x) w^{a \nu}(x) \nonu
\A \A \hspace{1.0cm}     
      =  g^{\mu\nu}  \nonu
\A \A \hspace{1.0cm}      
      - t^{\mu\nu} - t^{\nu\mu} \nonu
\A \A \hspace{1.0cm}     
      + t^{\rho\mu}{t^{\nu}}_{\rho} + t^{\rho\nu}{t^{\mu}}_{\rho} +  t^{\mu\rho}{t^{\nu}}_{\rho}  \nonu
\A \A \hspace{1.0cm}     
      - t^{\rho\mu}{t^{\sigma}}_{\rho} {t^{\nu}}_{\sigma} - t^{\rho\nu}{t^{\sigma}}_{\rho}{t^{\mu}}_{\rho}  
      - t^{\mu\sigma}{t^{\rho}}_{\sigma}{t^{\nu}}_{\rho} - t^{\nu\rho}{t^{\sigma}}_{\rho}{t^{\mu}}_{\sigma} \nonu
\A \A \hspace{1.0cm}      
      + t^{\rho\mu}{t^{\sigma}}_{\rho}{t^{\kappa}}_{\sigma}{t^{\nu}}_{\kappa} 
      + t^{\rho\nu}{t^{\sigma}}_{\rho}{t^{\kappa}}_{\sigma}{t^{\mu}}_{\kappa}         \nonu
\A \A \hspace{1.0cm}      
      + t^{\mu\sigma}{t^{\rho}}_{\sigma}{t^{\sigma}}_{\rho}{t^{\nu}}_{\sigma} 
      + t^{\nu\sigma}{t^{\rho}}_{\sigma}{t^{\sigma}}_{\rho}{t^{\mu}}_{\sigma} 
      + t^{\rho\kappa}{t^{\sigma}}_{\kappa} {t^{\mu}}_{\rho}{t^{\nu}}_{\sigma}. 
\label{new-si}
\ea
We can easily show 
\begin{equation}
%\eqalign{
{w_a}^{\mu} w_{b \mu} = \eta_{ab},   \quad    s_{\mu \nu}{w_a}^{\mu} {w_b}^{\nu} = \eta_{ab}.
%}
\label{new-metric}
\end{equation} 
It is obvious from the above general covariant arguments that (\ref{SGM}) is invariant 
under the ordinaly GL(4,R).     \\
By using (\ref{new-w}), (\ref{new-wi}), (\ref{new-s}) and (\ref{new-si}) we can express  
SGM action  (\ref{SGM}) in terms of  $e{^a}_{\mu}(x)$ and  $\psi^{j}(x)$, which describes explicitly 
the fundamental interaction of graviton with  superons. 
The expansion  of the action in terms of the power series of $\kappa$ (or ${t^{a}}_{\mu}$) 
can be carried out straightforwardly.  
After  the lengthy  calculations concerning the complicated  structures of the indices 
we obtain       \\   
\ba
\A \A L_{SGM} = - {c^3\Lambda \over 16{\pi}G} e \vert w_{V-A} \vert - {c^3 \over 16{\pi}G} e R  \nonu
\A \A \hspace{1.5cm}
+ {c^3 \over 16{\pi}G} e \big[ \ 2 t^{(\mu\nu)} R_{\mu\nu}  \nonu
\A \A \hspace{1.5cm}
+ {1 \over 2} \{ g^{\mu\nu}\partial^{\rho}\partial_{\rho}t_{(\mu\nu)}
- t_{(\mu\nu)}\partial^{\rho}\partial_{\rho}g^{\mu\nu}       \nonu
\A \A \hspace{1.5cm}
+ g^{\mu\nu}\partial^{\rho}t_{(\mu\sigma)}\partial^{\sigma}g_{\rho\nu}
- 2g^{\mu\nu}\partial^{\rho}t_{(\mu\nu)}\partial^{\sigma}g_{\rho\sigma}
- g^{\mu\nu}g^{\rho\sigma}\partial^{\kappa}t_{(\rho\sigma)}\partial^{\kappa}g_{\mu\nu} \}     \nonu
\A \A \hspace{1.5cm}
- 2({t^{\mu}}_{\rho}t^{\rho\nu}+{t^{\nu}}_{\rho}t^{\rho\mu}+t^{\mu\rho}{t^{\nu}}_{\rho})R_{\mu\nu}   \nonu
\A \A \hspace{1.5cm}
- \{t^{(\mu \rho)} t^{(\nu \sigma)} R_{\mu \nu \rho \sigma} \nonu
\A \A \hspace{1.5cm}
+ {1 \over 2}t^{(\mu\nu)}( g^{\rho\sigma}\partial^{\mu}\partial_{\nu}t_{(\rho\sigma)} 
- g^{\rho\sigma}\partial^{\rho}\partial_{\mu}t_{(\sigma\nu)} + \dots )  \}  \nonu
\A \A \hspace{1.5cm}
+\{ O(t^{3})\} +\{ O(t^{4})\} + \dots + \{ O(t^{10})\}) \big],
\label{L-exp}
\ea
where $e=det{e^{a}}_{\mu}$, $t^{(\mu\nu)}=t^{\mu\nu}+t^{\nu\mu}$, $t_{(\mu\nu)}=t_{\mu\nu}+t_{\nu\mu}$, and 
$ \vert w_{V-A} \vert = det{w^{a}}_{b} $ is the flat space V-A action\cite{va} 
containing up to $O(t^{4})$ and $R$ and $R_{\mu\nu}$ are the Ricci curvature tensors  of GR.    \\
Remarkably the first term can be regarded as a space-time dependent cosmological term
and  reduces to V-A action \cite{va} with ${\kappa_{V-A}}^{-1} = {c^3 \over 16{\pi}G}{\Lambda}$ 
in the Riemann-flat $e{_a}^{\mu}(x) \rightarrow \delta{_a}^{\mu}$ space-time. 
The second term is the familiar E-H action of GR. 
These expansions show the complementary relation of graviton and (the stress-energy tensor of) superons.  
The existence of (in the Riemann-flat space-time) NL SUSY invariant terms  
with the (second order) derivatives of the superons beyond V-A model are manifested. 
For example, the lowest order of such terms appear in $O(t^{2})$ and have the following expressions 
(up to the total derivative terms)                                           \\ 
\begin{equation}
%\eqalign{
-{c^3 \over {16 \pi G}} \epsilon^{abcd}{\epsilon_{a}}^{efg}\partial_{c}t_{(be)}\partial_{f}t_{(dg)}.
%}
\label{b-va}
\end{equation}
The existence of such  derivative  terms 
in addition to the original V-A model are already pointed out and exemplified  in part in \cite{sw}.  
Note that (\ref{b-va}) vanishes in 2 dimensional space-time.                                  \\
Here we just mention that we can consider two  types of the flat space in SGM, which are not equivalent. 
One is SGM-flat, i.e. ${w_{a}}^{\mu}(x) \rightarrow {\delta_{a}}^{\mu}$,  space-time and 
the other is Riemann-flat, i.e.  $e{_a}^{\mu}(x) \rightarrow \delta{_a}^{\mu}$, space-time, 
where SGM action reduces to  ${-{c^3\Lambda \over 16{\pi}G}}$ and                                
${-{c^3\Lambda \over 16{\pi}G} \vert w_{V-A} \vert - 
{c^3 \over 16{\pi}G}( derivative \  terms) }$, respectively. 
Note that SGM-flat space-time allows non trivial Riemann space-time.                \\  

{\bf 3. NLSUSY GR in two  dimensional(2D) space-time}    \\ 
It is well known that two dimensional GR has no physical degrees of freedom ( due to the local GL(2,R)). 
SGM in SGM space-time is also the case.
However the arguments with the general covariance shed light on the  characteristic off-shell gauge structures 
of the theory in any space-time dimensions. 
Especialy for SGM, it is also useful for linearlizing the theory  to see explicitly the superon-graviton 
coupling in (two dimensional) Riemann space-time.  The result gives the exact expansion  up to $O(t^{2})$ 
in four dimensional space-time as well.  \\
{\bf 3.1 2D NLSUSY GR in affine connection formalism}     \\
Now we go to two dimensional SGM space-time to simplify the arguments without loss of generality and 
demonstrate some details of the computations. 
We adopt firstly  the affine connection formalism. 
The knowledge of the complete  structure of SGM action including the surface terms 
is  useful to linearlize SGM into the equivalent linear theory and to find the symmetry breaking of the model.   \\
Following EGRT the scalar curvature tensor $\Omega$ of SGM space-time is given as follows 
\ba
\Omega \A = \A s^{\beta\mu}\Omega{^{\alpha}}_{\beta\mu \alpha}  \nonu
\A = \A s^{\beta\mu} [\{ \partial_{\mu}{\Gamma^{\lambda}}_{\beta\alpha} 
       +{\Gamma^{\alpha}}_{\lambda\mu}{\Gamma^{\lambda}}_{\beta\alpha} \} 
       -\{ \ lower \ indices  (\mu \leftrightarrow \alpha)  \}],      
%}
\label{Omega}
\ea
where the Christoffel symbol of the second kind of SGM space-time is 
\ba
\A \A {\Gamma^{\alpha}}_{\beta\mu}
={1 \over 2}s^{\alpha\rho}\{ \partial_{\beta}s_{\rho\mu}+\partial_{\mu}s_{\beta\rho}-\partial_{\rho}s_{\mu\beta}\}. 
%}
\label{affine}
\ea
The straightforwad expression of SGM action (\ref{SGM}) in two dimensional space-time, 
is given as follows     \\
\ba
\A \A L_{2dSGM} = - {c^3 \over 16{\pi}G} e \{ 1+{t^{a}}_{a}    
+{1 \over 2}({t^{a}}_{a}{t^{b}}_{b}-{t^{a}}_{b}{t^{b}}_{a}) \}  
(g^{\beta\mu}- {\tilde t}^{(\beta\mu)} + {\tilde t}^{2(\beta\mu)})    \nonu
\A \A \hspace{1.5cm}
\times [ \{ {1 \over 2}\partial_{\mu}(g^{\alpha\sigma}- \tilde t^{(\alpha\sigma)} + {\tilde t}^{2(\alpha\sigma)}) 
\partial_{\dot \beta}(g_{\dot \sigma \dot \alpha} +{{\b t}}_{(\dot\sigma \dot\alpha)}+
{{ {\b t}}^{2}}_{(\dot\sigma \dot\alpha)})    \nonu
\A \A \hspace{1.5cm}
+ {1 \over 2}(g^{\alpha\sigma}- \tilde t^{(\alpha\sigma)} + {\tilde t}^{2(\alpha\sigma)})
\partial_{\mu}\partial_{\dot \beta}
(g_{\dot \sigma \dot \alpha} + {\b t}_{(\dot\sigma \dot\alpha)}+{ {\b t}^{2}}_{(\dot\sigma \dot\alpha)} \}    \nonu
\A \A \hspace{1.5cm}
-\{ lower \ indices  (\mu \leftrightarrow \alpha) \}        \nonu
\A \A \hspace{1.5cm}
+\{ {1 \over 4}(g^{\alpha\sigma}- {\tilde t}^{(\alpha\sigma)} + {\tilde t}^{2(\alpha\sigma)})
\partial_{\dot \lambda}(g_{\dot\sigma \dot\mu} + {\b t}_{(\dot\sigma \dot\mu)}+
{{\b t}^{2}}_{(\dot\sigma \dot\mu)})  \nonu
\A \A \hspace{1.5cm} 
(g^{\lambda\rho}- {\tilde t}^{(\lambda\rho)} + {\tilde t}^{2(\lambda\rho)})      
\partial_{\dot \beta}(g_{\dot\rho \dot\alpha} + {\b t}_{(\dot\rho \dot\alpha)}+
{ {\b t}^{2}}_{(\dot\rho \dot\alpha)}) \}    \nonu
\A \A \hspace{1.5cm} 
- \{ lower \ indices  (\mu \leftrightarrow \alpha)  \}]   \nonu
\A \A \hspace{1.5cm}
- {c^3\Lambda \over 16{\pi}G} e \vert w_{V-A} \vert,     \nonu
\A \A \hspace{1.5cm}
%}
\label{2dSGM}
\ea
where we have put 
\ba
\A \A 
s_{\alpha\beta}=g_{\alpha\beta}+{\b t}_{(\alpha\beta)}
+{{\b t}^{2}}_{(\alpha\beta)}, 
\ \ s^{\alpha\beta}=g^{\alpha\beta}-{\tilde t}^{(\alpha\beta)}
+{\tilde t}^{2(\alpha\beta)},       \nonu
\A \A 
{\b t}_{(\mu\nu)}=t_{\mu\nu}+t_{\nu\mu}, 
\ \ {{\b t}^{2}}_{(\mu\nu)}={t^{\rho}}_{\mu}t_{\rho\nu},    \nonu
\A \A 
{\tilde t}^{(\mu\nu)}=t^{\mu\nu}+t^{\nu\mu},  
\ \ {\tilde t}^{2(\mu\nu)}={t^{\mu}}_{\rho}t^{\rho\nu}
+{t^{\nu}}_{\rho}t^{\rho\mu}+t^{\mu\rho}{t^{\nu}}_{\rho}, 
\label{2dSGM-not1}
\ea
and the Christoffel symbols of the first kind of SGM space-time 
contained in (\ref{affine}) are abbreviated as 
\ba
\A \A \partial_{\dot \mu}{ g}_{\dot \sigma \dot \nu} 
= \partial_{\mu}{g}_{\sigma\nu} 
+ \partial_{\nu}{g}_{\mu\sigma} 
- \partial_{\sigma}{ g}_{\nu\mu}, \nonu
\A \A 
\partial_{\dot \mu}{\b t}_{\dot \sigma \dot \nu} 
= \partial_{\mu}{\b t}_{(\sigma\nu)} 
+ \partial_{\nu}{\b t}_{(\mu\sigma)} 
- \partial_{\sigma}{\b t}_{(\nu\mu)}, \nonu
\A \A 
\partial_{\dot\mu}{{\b t}^{2}}_{\dot\sigma \dot\nu} 
= \partial_{\mu}{{\b t}^{2}}_{(\sigma\nu)} 
+ \partial_{\nu}{{\b t}^{2}}_{(\mu\sigma)} 
- \partial_{\sigma}{{\b t}^{2}}_{(\nu\mu)}. \nonu
\label{2dSGM-not2}
\ea
By  expanding the scalar curvature  $\Omega$ in the power series of $t$, 
we have the expression of two dimensional SGM (\ref{L-exp}) which terminates 
with $t^{4}$.   \\ 

{\bf 3.2 SGM in the spin connection formalism}       \\
Next we perform the similar arguments in the spin connection formalism 
for the sake of the comparison. 
The spin connection $Q{^{ab}}_{\mu}$ 
and the curvature tensor $\Omega{^{ab}}_{\mu \nu}$ 
in SGM space-time are as follows; 
\ba
\A \A Q_{ab \mu} 
= {1 \over 2} (w{_{[a}}^{\rho} \partial_{\mid \mu \mid} w_{b] \rho} 
- w{_{[a}}^{\rho} \partial_{\mid \rho \mid} w_{b] \mu} 
- w{_{[a}}^{\rho} w{_{b]}}^{\sigma} 
w_{c \mu} \partial_{\rho} w{^c}_{\sigma}), \\
\A \A \Omega{^{ab}}_{\mu \nu} 
= \partial_{[\mu} Q{^{ab}}_{\nu]} + Q{^a}_{c[\mu} Q{^{cb}}_{\nu]}. 
\ea
The scalar curvature $\Omega$ of SGM space-time is defined by 
$\Omega = w{_a}^{\mu} w{_b}^{\nu} \Omega{^{ab}}_{\mu \nu}$. 
Let us express the spin connection $Q{^{ab}}_{\mu}$ 
in two dimensional space-time in terms of 
$e{^a}_{\mu}$ and $t{^a}_{\mu}$ as 
\begin{equation}
Q_{ab \mu} = \omega_{ab \mu}[e] 
+ T^{(1)}_{ab \mu} + T^{(2)}_{ab \mu} 
+ T^{(3)}_{ab \mu}, 
\end{equation}
where $\omega_{ab \mu}[e]$ 
is the Ricci rotation coefficients of GR, 
and $T^{(1)}_{ab \mu}$, $T^{(2)}_{ab \mu}$ 
and $T^{(3)}_{ab \mu}$ are defined as 
\ba
T^{(1)}_{ab \mu} 
= \A \A {1 \over 2} (e{_{[a}}^{\rho} \partial_{\mid \mu \mid} t_{b] \rho} 
- t{^{\rho}}_{[a} \partial_{\mid \mu \mid} e_{b] \rho} 
- e{_{[a}}^{\rho} \partial_{\mid \rho \mid} t_{b] \mu} 
+ t{^{\rho}}_{[a} \partial_{\mid \rho \mid} e_{b] \mu} 
\nonu
\A \A - e{_{[a}}^{\rho} e{_{b]}}^{\sigma} 
e_{c \mu} \partial_{\rho} t{^c}_{\sigma} 
+ e{_{[a}}^{[\rho} t{^{\sigma]}}_{b]} 
e_{c \mu} \partial_{\rho} e{^c}_{\sigma} 
- e{_{[a}}^{\rho} e{_{b]}}^{\sigma} 
t_{c \mu} \partial_{\rho} e{^c}_{\sigma}), 
\\
T^{(2)}_{ab \mu} 
= \A \A {1 \over 2} (- t{^{\rho}}_{[a} \partial_{\mid \mu \mid} t_{b] \rho} 
+ t{^{\rho}}_{[a} t{^{\sigma}}_{\mid \rho} 
\partial_{\mu \mid} e_{b] \sigma} 
+ t{^{\rho}}_{[a} \partial_{\mid \rho \mid} t_{b] \mu} 
- t{^{\rho}}_{[a} t{^{\sigma}}_{\mid \rho} 
\partial_{\sigma \mid} e_{b] \mu} 
\nonu
\A \A + e{_{[a}}^{[\rho} t{^{\sigma]}}_{b]} 
e_{c \mu} \partial_{\rho} t{^c}_{\sigma} 
- e{_{[a}}^{\rho} e{_{b]}}^{\sigma} 
t_{c \mu} \partial_{\rho} t{^c}_{\sigma} 
- e{_{[a}}^{[\rho} t{^{\mid \sigma \mid}}_{b]} 
t{^{\lambda]}}_{\sigma} e_{c \mu} 
\partial_{\rho} e{^c}_{\lambda} 
\nonu
\A \A - t{^{\rho}}_{[a} t{^{\sigma}}_{b]} 
e_{c \mu} \partial_{\rho} e{^c}_{\sigma} 
+ e{_{[a}}^{[\rho} t{^{\sigma]}}_{b]} t_{c \mu} 
\partial_{\rho} e{^c}_{\sigma}), 
\\
T^{(3)}_{ab \mu} 
= \A \A {1 \over 2} (t{^{\rho}}_{[a} t{^{\sigma}}_{\mid \rho} 
\partial_{\mu \mid} t_{b] \sigma} 
- t{^{\rho}}_{[a} t{^{\sigma}}_{\mid \rho} 
\partial_{\sigma \mid} t_{b] \mu} 
\nonu
\A \A 
- e{_{[a}}^{[\rho} t{^{\mid \sigma \mid}}_{b]} 
t{^{\lambda]}}_{\sigma} e_{c \mu} 
\partial_{\rho} t{^c}_{\lambda} 
- t{^{\rho}}_{[a} t{^{\sigma}}_{b]} 
e_{c \mu} \partial_{\rho} t{^c}_{\sigma} 
+ e{_{[a}}^{[\rho} t{^{\sigma]}}_{b]} t_{c \mu} 
\partial_{\rho} t{^c}_{\sigma}), 
\ea
where $t{^{\mu}}_a = e{_b}^{\mu} e{_a}^{\nu} t{^b}_{\nu}$. 
Note that $T^{(1)}_{ab \mu}$ and $T^{(2)}_{ab \mu}$ 
can be written by using the spin connection 
$\omega{^{ab}}_{\mu}[e]$ of GR as 
\ba
T^{(1)}_{ab \mu} 
= \A \A e{_{[a}}^{\rho} \hat D_{\mid \mu \mid} t_{b] \rho} 
+ {1 \over 4} e{_{[a}}^{\rho} e{_{b]}}^{\sigma} 
\partial_{\dot\mu} t_{[\dot\rho \dot\sigma]}, 
\\
T^{(2)}_{ab \mu} 
= \A \A - t{^{\rho}}_{\sigma} 
e{_{[a}}^{\sigma} \hat D_{\mid \mu \mid} t_{b] \rho} 
+ {1 \over 2} e{_{[a}}^{\rho} e{_{b]}}^{\sigma} 
t_{c \rho} \hat D_{\mu} t{^c}_{\sigma} 
\nonu
\A \A 
- {1 \over 2} e{_{[a}}^{\rho} e{_{b]}}^{\sigma} 
\partial_{\rho} (t_{c \mu} t{^c}_{\sigma}) 
- {1 \over 2} t{^{\rho}}_{\lambda} 
e{_{[a}}^{\lambda} e{_{b]}}^{\sigma} 
\partial_{\dot\mu} t_{[\dot\rho \dot\sigma]}, 
\ea
where $\hat D_{\mu} t_{a \nu} := 
\partial_{\mu} t_{a \nu} 
+ \omega_{ab \mu} t{^{b}}_{\nu}$ 
and $\partial_{\dot\mu} t_{[\dot\rho \dot\sigma]} 
:= \partial_{\mu} t_{[\rho \sigma]} 
+ \partial_{\sigma} t_{(\mu \rho)} 
- \partial_{\rho} t_{(\sigma \mu)}$. 
Then we obtain straightforwardly 
the complete expression of 2 dimensional SGM action(N=1) 
in the spin connection formalism as follows; namely, up to $O(t^2)$  \\
\ba
L_{2dSGM} = \A \A 
- {{c^3 \Lambda} \over 16{\pi}G} e \vert w_{V-A} \vert 
\nonu
\A \A
- {c^3 \over 16{\pi}G} e \vert w_{V-A} \vert 
[R - 4 t^{\mu \nu} R_{\mu \nu} 
+ 2 e^{a[\mu} e^{\mid b \mid \nu]} 
(\hat D_{\mu} e{_a}^{\rho}) \hat D_{\nu} t_{b \rho} 
+ D_{\mu} (g^{\mu \rho} g^{\nu \sigma} 
\partial_{\dot\nu} t_{[\dot\rho \dot\sigma]}) 
\nonu
\A \A
+ 2 (t^{\rho \mu} t{^{\nu}}_{\rho} + t^{\rho \nu} t{^{\mu}}_{\rho} 
+ t^{\mu \rho} t{^{\nu}}_{\rho}) R_{\mu \nu} 
+ t^{(\mu \rho)} t^{(\nu \sigma)} R_{\mu \nu \rho \sigma} 
\nonu
\A \A
- (g^{\rho [\mu} g^{\mid \kappa \mid \nu]} g^{\sigma \lambda} 
+ g^{\sigma [\mu} g^{\mid \kappa \mid \nu]} g^{\rho \lambda} 
- g^{\sigma [\mu} g^{\mid \lambda \mid \nu]} g^{\kappa \rho}) 
e{^a}_{\sigma} e{^b}_{\kappa} (\hat D_{\mu} t_{a \rho}) 
\hat D_{\nu} t_{b \lambda} 
\nonu
\A \A
+ g^{\rho [\mu} g^{\mid \kappa \mid \nu]} g^{\sigma \lambda} 
e{^a}_{\sigma} (\hat D_{\mu} t_{a \rho}) 
\partial_{\dot\nu} t_{[\dot\lambda \dot\kappa]} 
+ {1 \over 4} g^{\rho [\mu} g^{\mid \kappa \mid \nu]} g^{\sigma \lambda} 
(\partial_{\dot\mu} t_{[\dot\rho \dot\sigma]}) 
\partial_{\dot\nu} t_{[\dot\lambda \dot\kappa]} 
\nonu
\A \A
- 2 (g^{\rho [\mu} e^{\mid b \mid \nu]} \hat D_{\mu} e^{a \sigma} 
+ e^{c [\mu} e^{\mid b \mid \nu]} e^{a \sigma} \hat D_{\mu} e{_c}^{\rho} 
\nonu
\A \A
+ e^{c [\mu} e^{\mid a \mid \nu]} e^{b \rho} \hat D_{\mu} e{_c}^{\sigma} 
- e^{b [\mu} e^{\mid a \mid \nu]} e^{c \rho} \hat D_{\mu} e{_c}^{\sigma}) 
t_{a \rho} \hat D_{\nu} t_{b \sigma} 
\nonu
\A \A
- e^{a [\mu} g^{\mid \rho \mid \nu]} (\hat D_{\mu} e{_a}^{\sigma}) 
t_{b [\rho} \hat D_{\mid \nu \mid} t{^b}_{\sigma]} 
- g^{\rho \nu} e^{a \mu} e^{b \sigma} (\hat D_{\mu} e{_b}^{\lambda}) 
t_{a \lambda} \partial_{\dot\nu} t_{[\dot\rho \dot\sigma]} 
\nonu
\A \A
- D_{\mu} \{ g^{\rho [\mu} g^{\mid \sigma \mid \nu]} 
\partial_{\rho}(t_{a \nu} t{^a}_{\sigma}) 
+ 2 g^{\rho [\mu} t^{\nu] \sigma} 
\partial_{\dot\nu} t_{[\dot\rho \dot\sigma]} \} 
+ \{ O(t^{3}) \} + \{ O(t^{4}) \}, 
\label{2dSGM-spin}
\ea
where $\hat D_{\mu} T_{a \nu} := 
\partial_{\mu} T_{a \nu} 
+ \omega_{ab \mu} T{^b}_{\nu}$ 
and $D_{\mu} e{_a}^{\nu} := \partial_{\mu} e{_a}^{\nu} 
+ \Gamma^{\nu}_{\lambda \mu} e{_a}^{\lambda}$.    \\

{\bf 4. Linearization of N=2 NLSUSY}             \\
The expansion of the SGM action in terms 
of graviton and superons (N-G fermions) with spin-1/2 
reveals a very complicated and rich structure; 
indeed, it is a highly nonlinear one which consists of 
the Einstein-Hilbert action of the general relativity, 
the V-A action and their interactions. 
Also, the SGM action is invariant under at least 
$[{\rm global\ nonlinear\ SUSY}] \otimes [{\rm local\ GL(4,R)}] 
\otimes [{\rm local\ Lorentz}] \otimes [{\rm global\ SO(N)}]$ 
as a whole, which is isomorphic to 
the global SO(N) super-Poincar\'e symmetry. 
\par
In the SGM the (composite) eigenstates of the {\it linear} 
representation of SO(10) super-Poincar\'e algebra which is composed 
of superons are regarded as all observed elementary 
particles at low energy except graviton \cite{ks1,ks2}. 
For deriving the low energy physical contents of the SGM action, 
it is important to linearize such a highly nonlinear theory and 
to obtain an equivalent renormalizable theory. 
In this respect the relationship between the $N = 1$ V-A model 
and a scalar supermultiplet of the linear SUSY  was 
well understood from the early work by many authors \cite{rikuzw}. 
\par
In this section we restrict our attention to the $N = 2$ SUSY 
and discuss a connection between the V-A model and an $N = 2$ 
vector supermultiplet  of the linear SUSY in 
four-dimensional spacetime\cite{stt2}. 
In particular, we show that for the 
$N = 2$ theory a SUSY invariant relation between component fields of 
the vector supermultiplet and the N-G fermion fields can be 
constructed by means of the method used in Ref.\ \cite{rikuzw} starting 
from an ansatz given below (Eq.\ (\ref{ansatz})). We also briefly 
discuss a relation of the actions for the two models. 
\par
Let us denote the component fields of an $N = 2$ U(1) gauge 
supermultiplet, which belong to representations of 
a rigid SU(2), as follows; 
namely, $\phi$ for a physical complex scalar field, 
$\lambda_R^i$ $(i = 1, 2)$ for two right-handed Weyl spinor fields 
and $A_a$ for a U(1) gauge field in addition to $D^I$ $(I = 1,2,3)$ 
for three auxiliary real scalar fields 
required from the mismatch of the off-shell degrees of freedom 
between bosonic and fermionic physical fields.\footnote{
Minkowski spacetime indices are denoted by $a, b, ... = 0, 1, 2, 3$, 
and the flat metric is $\eta^{ab}= {\rm diag}(+1, -1, -1, -1)$. 
Gamma matrices satisfy $\{ \gamma^a, \gamma^b \} = 2\eta^{ab}$ 
and we define $\gamma^{ab} = {1 \over 2}[\gamma^a, \gamma^b]$.}
$\lambda_R^i$ and $D^I$ belong to representations {\bf 2} and {\bf 3} 
of SU(2) respectively while other fields are singlets. 
By the charge conjugation we define left-handed spinor fields as 
$\lambda_{Li} = C \bar\lambda_{Ri}^T$. 
We use the antisymmetric symbols $\epsilon^{ij}$ and 
$\epsilon_{ij}$ ($\epsilon^{12} = \epsilon_{21} = +1$) to raise 
and lower SU(2) indices as $\psi^i = \epsilon^{ij} \psi_j$, 
$\psi_i = \epsilon_{ij} \psi^j$. 
\par
The $N=2$ linear SUSY transformations of these component fields 
generated by constant spinor parameters $\zeta_L^i$ are 
\ba
\delta_Q \phi \A = \A - \sqrt{2} \bar\zeta_R \lambda_L, \nonu
\delta_Q \lambda_{Li} =
\A \A - {1 \over 2} F_{ab} \gamma^{ab} \zeta_{Li} 
- \sqrt{2} i \partial \phi \zeta_{Ri} 
+ i ( \zeta_L \sigma^I )_i D^I, \nonu
\delta_Q A_a =
\A  \A - i \bar\zeta_L \gamma_a \lambda_L 
- i \bar\zeta_R \gamma_a \lambda_R, \nonu
\delta_Q D^I \A = \A \bar\zeta_L \sigma^I \partial \lambda_L 
+ \bar\zeta_R \sigma^I \partial \lambda_R, 
\label{lsusy}
\ea
where $\zeta_{Ri} = C \bar\zeta_{Li}^T$, 
$F_{ab} = \partial_a A_b - \partial_b A_a$, and $\sigma^I$ are 
the Pauli matrices. The contractions of SU(2) indices 
are defined as 
$\bar\zeta_R \lambda_L = \bar\zeta_{Ri} \lambda_L^i$, 
$\bar\zeta_R \sigma^I \lambda_L 
= \bar\zeta_{Ri} (\sigma^I)^i{}_j \lambda_L^j$, etc. 
These supertransformations satisfy a closed off-shell 
commutator algebra 
\ba
[ \delta_Q(\zeta_1), \delta_Q(\zeta_2)] 
= \delta_P(v) + \delta_g(\theta), 
\label{commutator}
\ea
where $\delta_P(v)$ and $\delta_g(\theta)$ are a translation and 
a U(1) gauge transformation with parameters 
\ba
v^a \A = \A 2i ( \bar\zeta_{1L} \gamma^a \zeta_{2L} 
- \bar\zeta_{1R} \gamma^a \zeta_{2R} ), \nonu
\theta \A = \A - v^a A_a + 2 \sqrt{2} \bar\zeta_{1L} \zeta_{2R} \phi 
- 2 \sqrt{2} \bar\zeta_{1R} \zeta_{2L} \phi^*. 
\label{u1}
\ea
Only the gauge field $A_a$ transforms under the U(1) gauge 
transformation 
\ba
\delta_g(\theta) A_a = \partial_a \theta. 
\ea
\par
Although our discussion on the relation between the linear and 
nonlinear SUSY transformations does not depend on a form of 
the action, it is instructive to consider 
a free action which is invariant under Eq.\ (\ref{lsusy}) 
\ba
S_{\rm lin} = \int d^4 x \left[ \partial_a \phi \partial^a \phi^* 
- {1 \over 4} F^2_{ab} 
+ i \bar\lambda_R \!\!\not\!\partial \lambda_R 
+ {1 \over 2} (D^I)^2 - {1 \over \kappa} \xi^I D^I \right], 
\label{lact}
\ea
where $\kappa$ is a constant whose dimension is $({\rm mass})^{-2}$ 
and $\xi^I$ are three arbitrary real parameters satisfying 
$(\xi^I)^2 = 1$. The last term proportional to $\kappa^{-1}$ is an 
analog of the Fayet-Iliopoulos $D$ term in the $N=1$ theories. 
The field equations for the auxiliary fields 
give $D^I = \xi^I / \kappa$ indicating a spontaneous SUSY breaking. 
\par
On the other hand, in the $N = 2$ V-A model \cite{BV} we have 
a nonlinear SUSY transformation law of the N-G fermion 
fields $\psi_L^i$ 
\ba
\delta_Q \psi_L^i = {1 \over \kappa} \zeta_L^i 
- i \kappa \left( \bar\zeta_L \gamma^a \psi_L 
- \bar\zeta_R \gamma^a \psi_R \right) 
\partial_a \psi_L^i, 
\label{nlsusy}
\ea
where $\psi_{Ri} = C \bar\psi_{Li}^T$. This transformation 
satisfies off-shell the commutator algebra (\ref{commutator}) 
without the U(1) gauge transformation on the right-hand side. 
The V-A action invariant under Eq.\ (\ref{nlsusy}) reads 
\ba
S_{\rm VA} = - {1 \over {2 \kappa^2}} \int d^4 x \; \det w, 
\label{vaact}
\ea
where the $4 \times 4$ matrix $w$ is defined by 
\ba
w{^a}_b = \delta^a_b + \kappa^2 t{^a}_b, \qquad 
t{^a}_b = - i \bar\psi_L \gamma^a \partial_b \psi_L 
+ i \bar\psi_R \gamma^a \partial_b \psi_R. 
\ea
The V-A action (\ref{vaact}) is expanded in $\kappa$ as 
\ba
S_{\rm VA} \A = \A 
- {1 \over {2 \kappa^2}} \int d^4 x 
\left[ 1 + \kappa^2 t{^a}_a 
+ {1 \over 2} \kappa^4 (t{^a}_a t{^b}_b 
- t{^a}_b t{^b}_a) \right. \nonu
\A\A
\left. - {1 \over 6} \kappa^6 \epsilon_{abcd} \epsilon^{efgd} 
t{^a}_e t{^b}_f t{^c}_g 
- {1 \over 4!} \kappa^8 \epsilon_{abcd} \epsilon^{efgh} 
t{^a}_e t{^b}_f t{^c}_g t{^d}_h 
\right]. 
\label{vaactex}
\ea
\par
We would like to obtain a SUSY invariant relation between the 
component fields of the $N = 2$ vector supermultiplet and 
the N-G fermion fields $\psi^i$ at the leading orders of $\kappa$. 
It is useful to imagine a situation in which the linear SUSY 
is broken with the auxiliary fields having expectation values 
$D^I = \xi^I / \kappa$ as in the free theory (\ref{lact}). 
Then, we expect from the experience in the $N = 1$ cases 
\cite{rikuzw}\cite{stt1}and the transformation law of the spinor 
fields in Eq.\ (\ref{lsusy}) that the relation should have a form 
\ba
\lambda_{Li} \A = \A i \xi^I (\psi_L \sigma^I)_i 
+ {\cal O}(\kappa^2), \nonu
D^I \A = \A {1 \over \kappa} \xi^I + {\cal O}(\kappa), \nonu
({\rm other\ fields}) \A = \A {\cal O}(\kappa). 
\label{ansatz}
\ea
Higher order terms are obtained such that the linear 
SUSY transformations (\ref{lsusy}) are reproduced by the nonlinear 
SUSY transformation of the N-G fermion fields (\ref{nlsusy}). 
\par
After some calculations we obtain the relation between the fields 
in the linear theory and the N-G fermion fields as 
\ba
\phi(\psi) \A = \A {1 \over \sqrt{2}} \, i \kappa \xi^I 
\bar\psi_R \sigma^I \psi_L 
- \sqrt{2} \kappa^3 \xi^I \bar\psi_L \gamma^a \psi_L 
\bar\psi_R \sigma^I \partial_a \psi_L \nonu
\A\A - {\sqrt{2} \over 3} \kappa^3 \xi^I \bar\psi_R \sigma^J \psi_L 
\bar\psi_R \sigma^J \sigma^I \!\!\not\!\partial \psi_R 
+ {\cal O}(\kappa^5), \nonu
\lambda_{Li}(\psi) \A = \A i \xi^I (\psi_L \sigma^I)_i 
+ \kappa^2 \xi^I \gamma^a \psi_{Ri} \bar\psi_R \sigma^I 
\partial_a \psi_L 
+ {1 \over 2} \kappa^2 \xi^I \gamma^{ab} \psi_{Li} 
\partial_a \left( \bar\psi_L \sigma^I \gamma_b \psi_L \right) \nonu
\A\A + {1 \over 2} \kappa^2 \xi^I ( \psi_L \sigma^J )_i 
\left( \bar\psi_L \sigma^J \sigma^I \!\!\not\!\partial \psi_L 
- \bar\psi_R \sigma^J \sigma^I \!\!\not\!\partial \psi_R \right) 
+ {\cal O}(\kappa^4), \nonu
A_a(\psi) \A = \A - {1 \over 2} \kappa \xi^I \left( 
\bar\psi_L \sigma^I \gamma_a \psi_L 
- \bar\psi_R \sigma^I \gamma_a \psi_R \right) \nonu
\A\A + {1 \over 4} i \kappa^3 \xi^I \biggl[ 
\bar\psi_L \sigma^J \psi_R \bar\psi_R \left( 2 \delta^{IJ} \delta_a^b 
- \sigma^J \sigma^I \gamma_a \gamma^b \right) \partial_b \psi_L \nonu
\A\A - {1 \over 4} \bar\psi_L \gamma^{cd} 
\psi_R \bar\psi_R \sigma^I \left( 2 \gamma_a \gamma_{cd} \gamma^b 
- \gamma^b \gamma_{cd} \gamma_a \right) \partial_b \psi_L 
+ (L \leftrightarrow R) \biggr] + {\cal O}(\kappa^5), \nonu
D^I(\psi) \A = \A {1 \over \kappa} \xi^I 
- i \kappa \xi^J \left( 
\bar\psi_L \sigma^I \sigma^J \!\!\not\!\partial \psi_L 
- \bar\psi_R \sigma^I \sigma^J \!\!\not\!\partial \psi_R \right) \nonu
\A\A + \kappa^3 \xi^J \biggl[ \bar\psi_L \sigma^I \psi_R \partial_a 
\bar\psi_R \sigma^J \partial^a \psi_L 
- \bar\psi_L \sigma^K \gamma^c \psi_L \biggl\{ 
i \epsilon^{IJK} \partial_c \bar\psi_L \!\!\not\!\partial \psi_L 
\nonu
\A\A - {1 \over 2} \partial_a \bar\psi_L 
\sigma^J \sigma^K \sigma^I \gamma_c \partial^a \psi_L 
+ {1 \over 4} \partial_a \bar\psi_L \sigma^J \sigma^I \sigma^K 
\gamma^a \gamma_c \!\!\not\!\partial \psi_L \biggr\} \nonu
\A\A - {1 \over 4} \bar\psi_L \sigma^K \psi_R \left\{ 
\partial_a \bar\psi_R \sigma^J \sigma^I \sigma^K \gamma^b \gamma^a 
\partial_b \psi_L 
- \bar\psi_R \left( 2 \delta^{IK} + \sigma^I \sigma^K \right) 
\sigma^J \Box \psi_L \right\} \nonu
\A\A + {1 \over 16} \bar\psi_L \gamma^{cd} \psi_R \left\{ 
\partial_a \bar\psi_R \sigma^J \sigma^I \gamma^b \gamma_{cd} 
\gamma^a \partial_b \psi_L 
+ \bar\psi_R \sigma^I \sigma^J \gamma^b \gamma_{cd} \gamma^a 
\partial_a \partial_b \psi_L \right\} \nonu
\A\A + (L \leftrightarrow R) \biggr] + {\cal O}(\kappa^5). 
\label{relation}
\ea
The transformation of the N-G fermion fields (\ref{nlsusy}) 
reproduces the transformation of the linear theory (\ref{lsusy}) 
except that the transformation of the gauge field 
$A_a(\psi)$ contains an extra U(1) gauge transformation 
\ba
\delta_Q A_a(\psi) = - i \bar\zeta_L \gamma_a \lambda_L(\psi) 
- i \bar\zeta_R \gamma_a \lambda_R(\psi) + \partial_a X, 
\ea
where 
\ba
X = {1 \over 2} i \kappa^2 \xi^I \bar\zeta_L \left( 
2 \delta^{IJ} - \sigma^{IJ} \right) \psi_R 
\bar\psi_R \sigma^J \psi_L + (L \leftrightarrow R). 
\ea
The U(1) gauge transformation parameter $X$ satisfies 
\ba
\delta_Q(\zeta_1) X(\zeta_2) 
- \delta_Q(\zeta_2) X(\zeta_1) = - \theta, 
\ea
where $\theta$ is defined in Eq.\ (\ref{u1}). 
Due to this extra term the commutator of two supertansformations 
on $A_a(\psi)$ does not contain the U(1) gauge transformation 
term in Eq.\ (\ref{commutator}). 
This should be the case since the commutator on $\psi$ does not 
contain the U(1) gauge transformation term. 
For gauge invariant quantities like $F_{ab}$ the transformations 
exactly coincide with those of the linear SUSY. 
In principle we can continue to obtain higher order terms in the 
relation (\ref{relation}) following this approach. 
However, it will be more useful to use the $N=2$ superfield 
formalism  as was done in Refs.\ \cite{rikuzw} 
for the $N=1$ theories. 
\par
We note that the leading terms of $A_a$ in Eq.\ (\ref{relation}) 
can be written as 
\ba
A_a = - \kappa \xi^1 \bar\chi \gamma_5 \gamma_a \varphi 
+ i \kappa \xi^2 \bar\chi \gamma_a \varphi 
- {1 \over 2} \kappa \xi^3 \left( \bar\chi \gamma_5 \gamma_a \chi 
- \bar\varphi \gamma_5 \gamma_a \varphi \right) 
+ {\cal O}(\kappa^3), 
\ea
where we have defined Majorana spinor fields 
\ba
\chi = \psi_L^1 + \psi_{R1}, \qquad 
\varphi = \psi_L^2 + \psi_{R2}. 
\ea
When $\xi^1 = \xi^3 = 0$, this shows the vector nature of the U(1) 
gauge field as we expected. 
\par
The relation (\ref{relation}) reduces to that of the $N = 1$ 
SUSY by imposing, e.g.  $\psi_L^2 = 0$. 
When $\xi^1 = 1$, $\xi^2 = \xi^3 = 0$, we find 
$\lambda_{L2} = 0$, $A_a = 0$, $D^3 = 0$ and that the relation 
between $(\phi, \lambda_{L1}, D^1, D^2)$ and $\psi_L^1$ 
becomes that of the $N=1$ scalar supermultiplet obtained in 
Ref.\ \cite{rikuzw}. 
When $\xi^1 = \xi^2 = 0$, $\xi^3 = 1$, on the other hand, 
we find $\lambda_{L1} = 0$, $\phi = 0$, $D^1 = D^2 = 0$ 
and that the relation between $(\lambda_{L2}, A_a, D^3)$ and 
$\psi_L^1$ becomes that of the $N=1$ vector 
supermultiplet obtained in Refs.\ \cite{rikuzw,stt1}. 
\par
Our result (\ref{relation}) does not depend on a form of the 
action for the linear SUSY theory. 
We discuss here the relation between the free linear SUSY action 
$S_{\rm lin}$ in Eq.\ (\ref{lact}) and the V-A action 
$S_{\rm VA}$ in Eq.\ (\ref{vaact}). 
It is expected that they coincide when Eq.\ (\ref{relation}) is 
substituted into the linear action (\ref{lact}) as in the 
$N=1$ case \cite{rikuzw,stt1}. 
We have explicitly shown that $S_{\rm lin}$ indeed coincides 
with the V-A action $S_{\rm VA}$ up to and including 
$O(\kappa^0)$ in Eq.\ (\ref{vaactex}). 
\par
Finally we summarize our results. In this paper we have constructed 
the SUSY invariant relation between the component fields 
of the $N = 2$ vector supermultiplet and the N-G fermion fields 
$\psi_L^i$ at the leading orders of $\kappa$. 
We have explicitly showed that the U(1) gauge field $A_a$ 
has the vector nature in terms of the N-G fermion fields 
in contrast to the models with the $N = 1$ SUSY \cite{stt1}. 
The relation (\ref{relation}) contains three arbitrary real 
parameters $\xi^I/\kappa$, which can be regarded as the vacuum 
expectation values of the auxiliary fields $D^I$. 
When we put $\psi_L^2 = 0$, the relation reduces to that of 
the $N = 1$ scalar supermultiplet or that of the $N = 1$ vector 
supermultiplet depending on the choice of the parameters $\xi^I$. 
We have also shown that the free action $S_{\rm lin}$ 
in Eq.\ (\ref{lact}) with the Fayet-Iliopoulos $D$ term 
reduces to the V-A action $S_{\rm VA}$ in Eq.\ (\ref{vaact}) 
at least up to and including $O(\kappa^0)$. 
From the results in this section we anticipate 
the equivalence of the action of $N$-extended standard 
supermultiplets to the $N$-extended V-A action 
of a nonlinear SUSY.   \
A U(1) gauge field, though an axial vector field for N=1 case, is expressed by N-G field 
(and its highly  nonlinear self interactions).  
The linearlization of N=2 V-A model is very important from the physical point of view, 
for it gives a new mechanism  which  generates a U(1) gauge field of the linearlized (effective) 
theory \cite{ks4}. 
In fact, as for the linearlization of N=2 V-A model, we have shown  that  
a realistic U(1) gauge field, i.e. a vector gauge field,  can be obtained by the systematic linearlization 
\cite {stt2}. 
It is remarkable that the renormalizable field theoretical model 
is obtained systematically by the linearlization of V-A model. 
In our case of SGM, the algebra( gauge symmetry ) should be changed by the linearlization 
from (13) to broken SO(10) SP(broken SUGRA \cite{FVF}\cite{DZ})symmetry, 
which are isomorphic. 
The systematic and generic arguments on the relation of 
linear and nonlinear SUSY are already investigated\cite{wb}. 
All these arguments are the encouraging and favourable results 
towards the linearlization of SGM. 
Indeed, we have recently discussed on some systematics 
in the linearization of SGM for $N = 1$ SUSY 
in the superspace formalism \cite{st3}.   \\

{\bf 5. Discussions}             \\
We have shown explicitly that contrary 
to its simple expression (\ref{SGM}) 
in unified SGM space-time the  expansion of SGM action shows 
very complicated and rich structures 
describing as a whole the gauge invariant graviton-superon interactions. 
The explicit expression of the expansion of SGM action is useful 
for determining the structure 
of the linear theory which is equivalent to SGM.          \\
The linearization of NLSUSY GRT(N=1 SGM) action (\ref{SGM}), i.e. the construction (identification) of 
the renormalizable and local LSUSY {\it gauge} field theory which is equivalent to (\ref{SGM}), 
is inevitable to derive the SM as the low energy effective theory of N=1 SGM. 
Particularly N=10 must be linearized to test the composite SGM scenario, 
though some characterictic and accessible predictions, e.g. 
nutrino- and  quark-mxings, proton stability, CP violation, the generation structure.. etc. 
are obtained qualitatively by the group theoretical arguments\cite{ks0}\cite{ks2}. 
The linearizations of  N=1 and N=2  NLSUSY VA action in flat spacetime  have been carried out explicitly 
by the systematic arguments and show  that they are equivalent to the LSUSY actions 
with Fayet-Iliopoulos terms for the scalar (or axial vector) supermultiplet\cite{rikuzw}
and the vector supermultiplet\cite{stt1}, respectively. 
These exact results obtained systematically as the representations of the symmetries  
by {\em the algebraic arguments} are favourable and encoraging towards 
the linearization of SGM and the (composite) SGM scenario. 
We anticipate that the local spacetime  symmetries  of SGM mentioned above  plays  crucial roles 
in the linearization, especially in constructing the SUSY invariant relations\cite{rikuzw}.     \\
We regard that the ultimate real shape of nature is  high symmetric new(SGM) spacetime inspired by NLSUSY, 
where the coset space coordinates $\psi$ of ${superGL(4,R) \over GL(4,R)}$ 
turning to the NG fermion d.o.f. in addition to the ordinaly Minkowski coordinate $x^{a}$, 
i.e.  $ local \ SL(2C) \times local \ SO(3,1)$ d.o.f., are attached at every  spacetime point.  
The geometry of new spacetime is described by SGM action (\ref{SGM}) of {\it vacuum \ EH-type} 
and gives the unified description of nature. 
As proved for EH action of GR\cite{wttn}, the energy of NLSUSY GR action of 
EH-type is anticipated to be positive ( $\Lambda>0$).  
NLSUSY GR action (\ref{SGM}), $L(w) \sim w\Omega + w\Lambda$, on SGM spacetime is unstable  and 
induces {\it the spontaneous (symmetry) breakdown} into EH action 
with NG fermion (massless superon-quintet) matter, 
${L(e,\psi) \sim eR + e\Lambda + (\cdots \kappa,\psi \cdots)}$\cite{st1}, on ordinary Riemann spacetime, 
for the curvature-energy potential of SGM  spacetime is released into the potential of Riemann spacetime 
and the energy-momentum of superon(matter), i.e. ${w\Omega > eR}$. 
As mentioned before SGM action poseses  two different flat spaces. 
One is SGM-flat ($w{^a}_{\mu}(x) \rightarrow {\delta}{^a}_{\mu}$)  space of NLSUSY GR action $L(w)$. 
And the other is Riemann-flat ($e{^a}_{\mu}(x) \rightarrow {\delta}{^a}_{\mu}$) space of SGM action ${L(e,\psi)}$ 
which allows (generalized) NLSUSY VA action. 
This can be regarded as the phase transition of spacetime from SGM to Riemann (with NG fermion matter).
Also this may be the birth of the present expanding universe, i.e. the big bang and 
the rapid expansion (inflation) of spacetime and matter, followed by the present observed 
expansion due to the small cosmological constant $\Lambda$.  
And we think that the birth of the present universe by the {\it spontaneous \ breakdown} 
of SGM spacetime  described  by {\it vacuum} action of EH-type (\ref{SGM}) 
may explain qualitatively the observed critical value$(\sim 1)$  of the energy density 
of the universe and (the absence of) the gravitational catastrophe.  
It is interesting if SGM could give new insights into the unsolved problems 
of the cosmology, e.g. the origin(real shape) of the  big bang, inflation(inflaton field), dark energy,
matter-antimatter asymmetry, $\cdots$, etc.   \par
In this study we have attempted a {\it geometrical} unification of spacetime and matter.
New (SGM) spacetime is the ultimate physical entity and specified by NLSUSY GRT (SGM action) (\ref{SGM})  
of vacuum EH-type. 
The study of the vacuum structure of SGM action in the broken phase
(i.e. NLSUSY GRT action in Riemann spacetime with matter) is important 
for linearizing SGM and to obtain the equivalent local LSUSY gauge field theory.            \par
SGM  with the extra dimensions, which  can be constructed straightforwardly and 
gives another unification framework by regarding the observed particles as elementary, is open.  
In this case there are two mechanisms for the conversion of the spacetime d.o.f. 
into the dynamical d.o.f., i.e. by the compactification of Kaluza-Klein type and 
by the new mechanism presented in SGM.  \\
Besides the composite picture of SGM it is interesting to consider (elementary field) SGM with the extra dimensions 
and their  compactifications. The compactification of ${w^{A}}_M={e^{A}}_M + {t^{A}}_M, (A,M=0,1,..D-1)$  
produces rich spectrum of bosons and fermions, which may give a new framework for 
the unification of space-time and matter.          \\
Also SGM for spin ${3 \over 2}$ superon(N-G fermion)\cite{st1} is formally within the same scope.
The cosmology of NLSUSY GR is open.   
\newpage

%%%%%%%  References  %%%%%%%%%%%%%%%%%%%%%%%%%%%%%%%%%%%%%%%
%
\newcommand{\NP}[1]{{\it Nucl.\ Phys.\ }{\bf #1}}
\newcommand{\PL}[1]{{\it Phys.\ Lett.\ }{\bf #1}}
\newcommand{\CMP}[1]{{\it Commun.\ Math.\ Phys.\ }{\bf #1}}
\newcommand{\MPL}[1]{{\it Mod.\ Phys.\ Lett.\ }{\bf #1}}
\newcommand{\IJMP}[1]{{\it Int.\ J. Mod.\ Phys.\ }{\bf #1}}
\newcommand{\PR}[1]{{\it Phys.\ Rev.\ }{\bf #1}}
\newcommand{\PRL}[1]{{\it Phys.\ Rev.\ Lett.\ }{\bf #1}}
\newcommand{\PTP}[1]{{\it Prog.\ Theor.\ Phys.\ }{\bf #1}}
\newcommand{\PTPS}[1]{{\it Prog.\ Theor.\ Phys.\ Suppl.\ }{\bf #1}}
\newcommand{\AP}[1]{{\it Ann.\ Phys.\ }{\bf #1}}

\end{document}